\documentstyle[aps,multicol,eqsecnum,epsf]{revtex}

\begin{document}
\draft

\title{Dynamics of Granular Stratification}

\author{Hern\'an A. Makse,$^{1,2}$ Robin C. Ball,$^1$ 
H. Eugene Stanley$^2$, and Stephen Warr$^1$}

\address{$^1$ Cavendish Laboratory, University of Cambridge, Madingley
Road, Cambridge CB3 0HE, UK\\
$^2$ Center for Polymer Studies and Physics Dept., Boston
University, Boston, MA 02215 USA}

\date{Phys. Rev. E {\bf 58}, 3357 (1998)}

\maketitle
\begin{abstract}

Spontaneous stratification in granular mixtures---i.e. the formation of
alternating layers of small-rounded and large-faceted grains when one
pours a random mixture of the two types of grains into a quasi-two
dimensional vertical Hele-Shaw cell---has been recently reported by H.
A. Makse et al. [Nature {\bf 386}, 379 (1997)].   Here we study
experimentally the dynamical processes leading to spontaneous
stratification.   We
divide the process in three stages: (a) avalanche of grains and
segregation in the rolling phase, (b) formation of the ``kink''---an
uphill wave at which grains are stopped---at the bottom substrate, and
(c) uphill motion of the kink and formation of a pair of layers.  Using
a high-speed video camera, we study a rapid flow regime where the
rolling grains size segregate during the avalanche due to the fact that
small grains move downward in the rolling phase to form a sublayer of
small rolling grains underneath a sublayer of large rolling grains.
This dynamical segregation process---known as 
 ``kinematic sieving'', ``free surface segregation'' or simple 
``percolation''--- contributes to the spontaneous
stratification of grains in the case of thick flows. 
We characterize the dynamical process of
stratification by measuring all relevant quantities: the velocity of the
rolling grains, the velocity of the kink, and the wavelength of the
layers. We also measure other phenomenological constants such as the
rate of collision between rolling and static grains, and all the angles
of repose characterizing the mixture. The wavelength of the layers
behaves linearly with the thickness of the layer of rolling grains
(i.e., with the flow rate), in agreement with theoretical predictions.
The velocity profile of the grains in the rolling phase is a linear
function of the position of the grains along the moving layer, which
implies a linear relation between the mean velocity and the thickness of
the rolling phase. We also find that the speed of the upward-moving kink
has the same value as the mean speed of the downward-moving grains. We
measure the shape and size of the kink, as well as the profiles of the
rolling and static phases of grains, and find agreement with recent
theoretical predictions.
\end{abstract}

\begin{multicols}{2}                            

\narrowtext

\section{Introduction}
\label{introduction}

Size segregation of granular mixtures
\cite{bagnold2,borges,nagel,bideaux,mehta,wolf,varenna} is known to
occur when mixtures are exposed to external periodic perturbations. A
much-studied size segregation phenomenon is known as the ``Brazil nut
effect'' \cite{williams,rosato,herrmann,knight,warr} and occurs when,
upon vibration, larger grains rise on a bed of finer grains. Axial size
segregation in alternating bands consisting of small and large grains
occurs when a mixture of grains is placed in a horizontal rotating
cylinder \cite{zik,kaka1,duran,cantalupe}. It is also known that even in
the absence of external perturbations mixtures of grains of different
sizes can spontaneously segregate. For example, when a mixture of
spherical grains of different sizes is poured onto a heap, the large
grains are more likely to be found near the base, while the small grains
are more likely to be near the top
\cite{brown,bagnold0,williams63,drahun,drahun2,fayed,savage1,savage2,%
savage3,meakin}.

\begin{figure}
\centerline{
\vspace*{.5cm}
\vbox{{\bf (a)}
 \epsfxsize=8cm \epsfbox{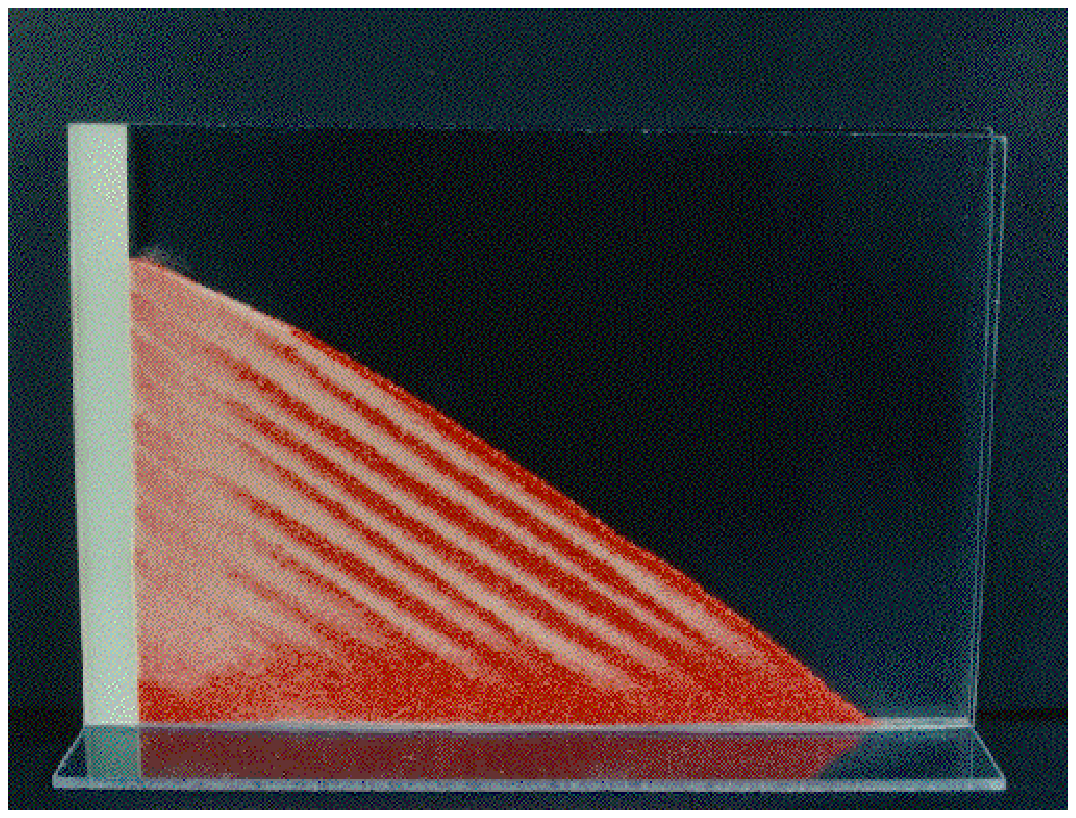} }
    }
\vspace*{.5cm}
\centerline{
\vbox{ {\bf (b)}
\epsfxsize=8cm \epsfbox{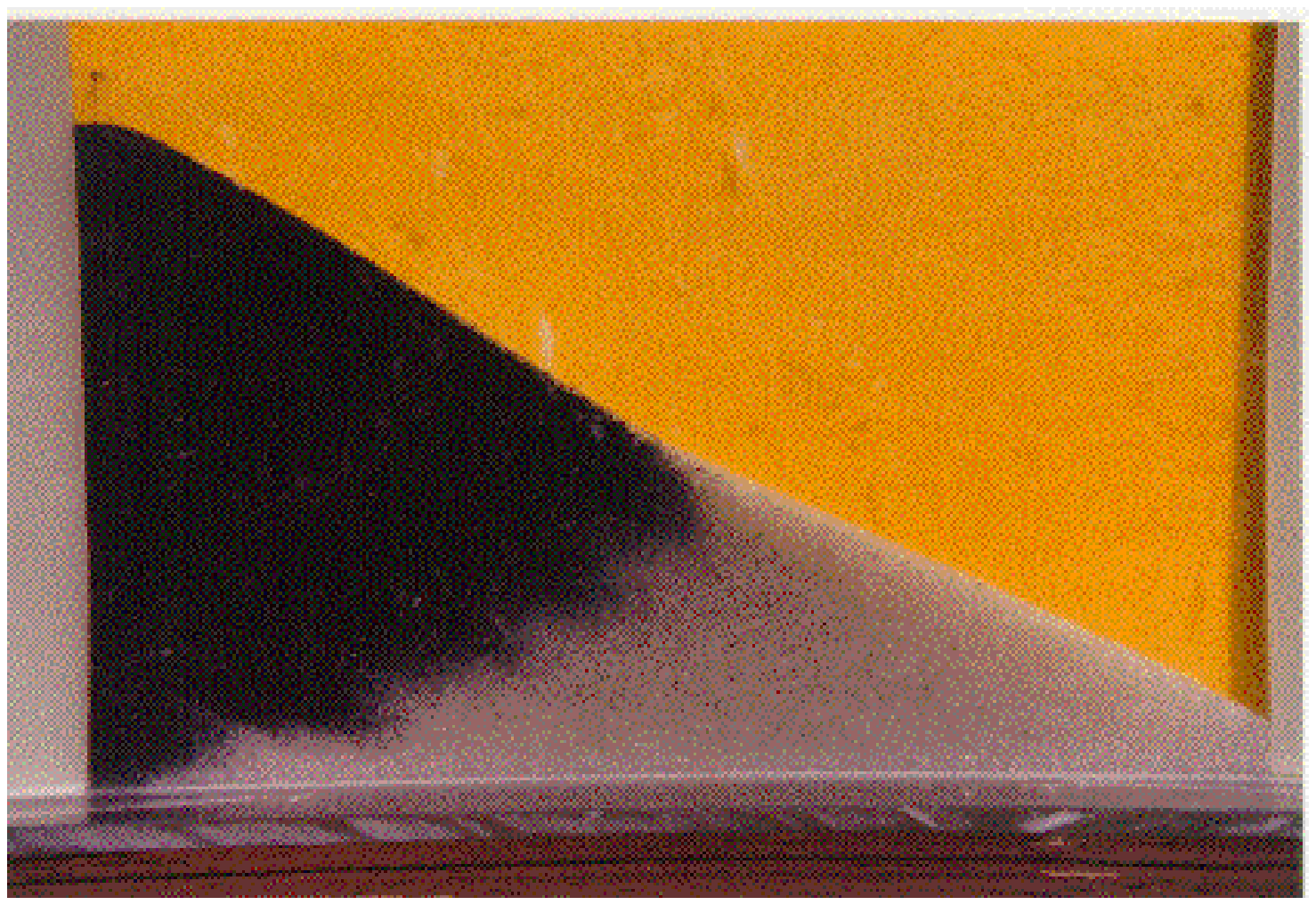 }}
   } 
\narrowtext
\caption{(a) Example of  stratification for a mixture
of smaller rounded grains (white spherical glass beads) and larger
faceted grains (black sugar grains). (b) Example of sharp 
segregation for a
mixture of smaller faceted grains (black sand) and larger rounded grains
(white spherical glass beads). Notice the sharp 
zone of separation of the species in the middle 
of the pile. This is the result of strong segregation effects acting 
in the system. Notice also the smaller angle of repose of the spherical beads
at the bottom of the pile.}
\label{strat}
\end{figure}

Another type of segregation, called spontaneous
stratification, arises when the
grains composing the mixture differ not only in size but also in shape
(or friction properties).  When a mixture of large grains that are more
faceted and small grains that are less faceted is poured in a ``granular
Hele-Shaw cell'' (two vertical slabs separated by a gap of typically
5--10 mm), the mixture spontaneously stratifies into alternating layers
of larger faceted grains and smaller rounded grains \cite{makse}. Figure
\ref{strat}a shows an example of such stratification. A mixture of large
cubic sugar grains (typical diameter 0.8 mm) and smaller spherical glass
beads (diameter 0.19 mm) is poured in the cell. We notice the striped
pattern with approximately constant wavelength.

In contrast, when the mixture is composed of larger less faceted grains
and smaller more faceted grains, the mixture only segregates---i.e., the
small more-faceted grains are found preferentially at the top of the
cell, while the large less-faceted grains are found near the bottom
\cite{makse}. Figure \ref{strat}b shows an example of such segregation, 
when a mixture of small faceted sand grains (typical size 0.3 mm) and
large spherical glass beads (typical size 0.8 mm) is poured in the cell.

The dynamical process leading to spontaneous
stratification was recently studied
theoretically \cite{makse2,makse3,cms,makse4} using discrete models, and
a set of continuum equations for surfaces flows of granular mixtures
developed in Refs.~\cite{bouchaud,bouchaud2,pgg,bdg}. In this
theoretical formalism, the grains are considered to belong to one of two
phases: a {\it static or bulk phase\/} if the grain is part of the solid
sandpile, and a {\it rolling or liquid phase } if the grain is not part
of the sandpile but rolls downward on top of the static phase. In
Ref.~\cite{makse2} the dynamics of spontaneous stratification was found
to be governed by the existence of a ``kink'' at which the grains are
stopped during an avalanche.

\begin{figure}
\centerline{
\vspace*{0.5cm}
\vbox{ \epsfxsize=8cm \epsfbox{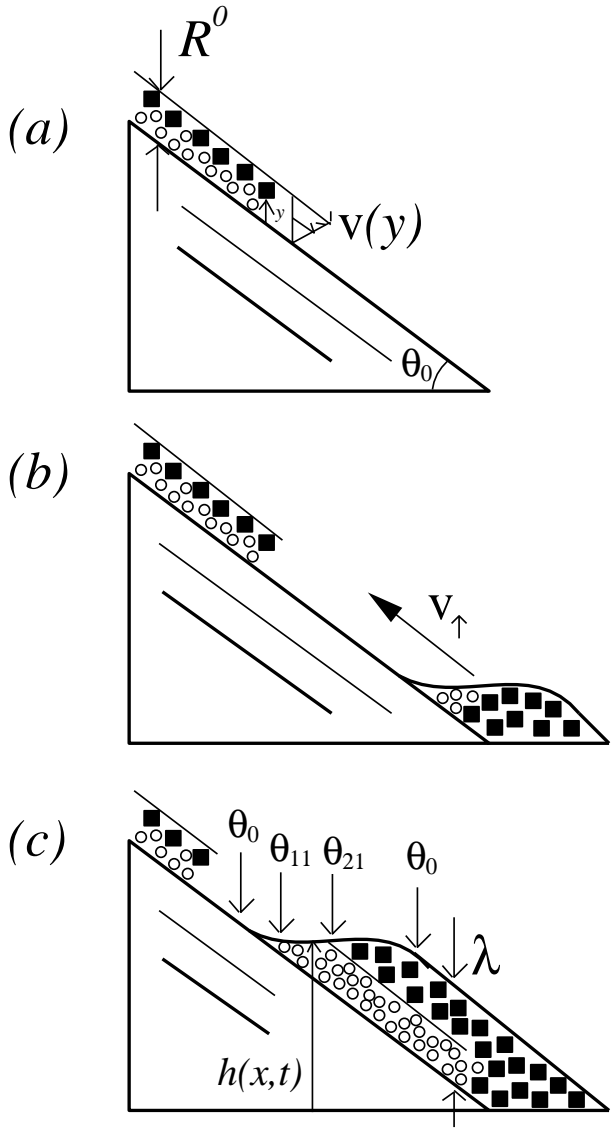} }
    }
\narrowtext
\caption{Three stages of the dynamics of  stratification:
(a) avalanche of grains, (b) formation of the kink at the bottom
substrate, (c) uphill motion of the kink. The dashed line in (b) is the window of observation used to 
record the images.}
\label{stages}
\end{figure}

In this paper, we study experimentally the dynamical processes leading
to spontaneous
stratification. Using a high speed video camera to study the motion
of the grains in great detail, we divide the dynamical process of
stratification into three stages (see Fig. \ref{stages}):
\begin{itemize}
\item[(a)] The avalanche of grains down the slope, and size 
segregation of grains in the rolling phase due to 
``percolation''.
\item[(b)] 
The formation of the ``kink''---an uphill wave at which grains are
stopped.
\item[(c)] The uphill motion of the kink and formation of a pair of layers.
\end{itemize}

We study a well-developed flow regime where the rolling grains segregate
during the flow. In this regime the thickness of the layer of rolling
grains is larger than the typical size of a grain $d$ (typically 5$d$),
and the smaller rolling grains are found to percolate downward in the
rolling phase to form a sublayer of smaller rolling grains underneath
the sublayer of larger rolling grains. This dynamical size segregation
process, known as ``percolation'' or ``kinematic sieving''
\cite{drahun,drahun2,savage1,savage2,savage3},
contributes to the stratification of grains.

Stratification is an instability developed due to a competition between
size segregation and shape segregation \cite{makse3}. 
In the case of thin flows,
size segregation occurs since the smaller grains are
captured more easily than larger grains. In the case of thick flow
regimes study here, the kinematic sieving in the rolling phase is mainly
responsible for the size segregation of the grains. Since the larger
grains are on top of the rolling phase, they are convected further down
than the smaller grains, producing the size segregation effect, which
together with the segregation due to different shape of the grains,
gives rise to the instability leading the system to spontaneously
stratify \cite{makse3}.
It is important to note that percolation in the rolling phase
is not sufficient condition to obtain stratification.
For thick flows and when the large grains are smoother, segregation in 
the rolling phase still occurs, and yet we do not get stratification but
only the sharp segregation pattern of Fig. \ref{strat}b.

A large difference in size is also a condition for the percolation 
effect to take place--- 
usually $\rho>1.5$, where $\rho$ is the ratio of the 
size of the large grains to the size of the 
small grains. We performed 
a series of experiments 
with mixtures of glass beads and sand with $\rho<1.5$  and found 
continuos segregation patterns 
(as opposed to the sharp segregation pattern
with a 
separation zone of a few centimeters 
of Fig. \ref{strat}b obtained for $\rho>1.5$) 
no matter the shape of the grains. This is because, when $\rho<1.5$ the 
effect of size segregation is very weak.

The limiting case in which both species of grains are spherical was
first studied by Williams
\cite{williams63,williams68,allen}; his results (showing
segregation plus a hint of stratification) differ from our results for
this case (showing only segregation).
We believe that the reason is that
the grains used by Williams were not quite spherical, inducing some shape 
segregation as well.
According to the above interpretation, we note that experiments 
with  mixtures of
perfect spherical beads differing only in size
should not show
stratification since the shape segregation effect is not present and size 
segregation alone (even due to percolation) is not able to produce 
stratification--- and our work confirms these expectations.  However some
oscillations might still be present around the stable segregation
profile, as seen in previous experiments using mixtures of spherical
beads \cite{allen}.

Here we focus on the regime where segregation in the rolling phase takes 
place.
We characterize the kinematic percolation process, and measure the
velocity gradient of the grains during the avalanche. We find a linear
velocity profile of the rolling grains, and that the mean velocity of
the rolling grains is the same as the velocity of the kink. We also
study the shape and size of the kink---and thereby measure the
wavelength of the layers. We find that the wavelength increases linearly
with the flux of grains, a result in agreement with recent theoretical
predictions \cite{makse2}. We also measure the profiles of the rolling
grains and static phases of the pile, and the values of several
phenomenological coefficients which appear in the theory for surface
flows of granular mixtures.  Our results are valid for flow rates of the
order of gr/sec (which gives rise to a rolling phase less than 1 cm
thick).  We also comment on the applicability of our results, and on the
deviations that may occur for smaller and larger flow regimes.

\section{Experimental setup}
\label{sectionsetup}

Our experimental setup consists of a granular Hele-Shaw cell: a vertical
``quasi-two-dimensional'' cell with a narrow gap separating two
transparent plates (made of plexiglass, or of glass). The cell measures
$L=30$ cm of lateral size and $20$ cm high, and the gap is $\ell=0.5$
cm. We close the left edge of the cell. We clean the walls of the cell
with an antistatic cleaner in order to avoid the effects of
electrostatic interaction between the grains and the wall.

In this study, we focus on spontaneous 
stratification. In all our experiments, we
use a mixture of grains composed of two species differing in size and
shape: smaller glass beads of average diameter $0.19\pm0.05$ mm,
spherical shape ($95\%$ sphericity), angle of repose
$\theta_{11}=26^\circ\pm 1^\circ$ (we call these type 1 grains), and
larger faceted sugar grains of typical size $0.8$ mm, approximate cubic
shape and angle of repose $\theta_{22}=39^\circ\pm 1^\circ$ (type 2
grains).

The typical size of the sugar grains obtained by measuring the volume of
the cubic grains and calculating the typical size as the cubic root
averaged over 20 different grains. We obtain the value of the angle of
repose of the species by pouring the pure species in the Hele-Shaw cell
and measuring the resulting angle of the pile, averaging over 5
realizations of the sandpile. The angle thus measured is not the actual
angle of repose corresponding to a conical pile, since the presence of
the wall induces extra friction that slightly increases the equilibrium
angle of the pile \cite{herrmann2}. However, we are interested in the
angle of repose for this specific geometry since our experiments on
stratification are done in the cell.

We fill the cell at different rates of adding grains (flux). We use a
Kodak Ektapro 1000 digital high speed camera to film the motion of
individual grains during the formation of the layers. The camera
produces 1000 digital frames per second with a resolution of 238
$\times$ 191 pixels.  We record images during 1.6 sec, and achieve
longer recording times by lowering the frame rate.  The digital images
we download to a workstation for further image processing.

\section{The angle of repose of the pure species}

\begin{figure}
\centerline{
\vspace*{0.5cm}
\vbox{ \epsfxsize=6cm \epsfbox{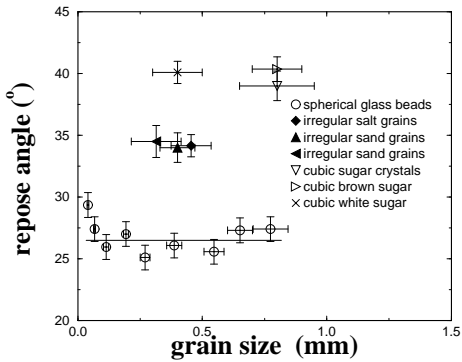}}
    }
\narrowtext
\caption{Angle of repose of different set of spherical glass beads
differing only in size, and other species such as sand, salt and sugar.
The more irregular is the shape of the grains, the larger is the angle
of repose.}
\label{repose}
\end{figure}

Since stratification is related to the different angles of repose of the
pure species, we first study how the angle of repose depends on the size
and shape of the grains. We measure the angle of repose of different
sets of spherical glass beads of different size, and find the same value
within errors bars (see Fig.
\ref{repose}). The angle of repose does not depend on the size of the
grains, since a simple isotropic rescaling of the pile coordinates
transforms a pile of smaller spherical grains into a pile of large
spherical grains, while leaving invariant the angle of the pile.

The value of the angle of repose we find for the {\it spherical\/} glass
beads is smaller than the value we find for the {\it cubic\/} sugar
grains. In general, we find that the angle of repose does not depend on
the size of the grains, and is a function of the shape of the grains:
the rougher the shape of the grains the larger the angle of
repose---because for more faceted grains the packing of grains is less
dense than for more rounded grains.

A particular case is found when the size of the grains is of the order
of microns. Spherical beads of size $40 \mu$m have a larger angle of
repose than the same spherical beads but of size of the order of mm (see
Fig. \ref{repose})---because adhesion forces become important,
increasing the angle of repose of the species.  The scale of microns is
the lower limit of applicability of our results, 
since at the submicron scale particles undergo Brownian
motion \cite{varenna} and our analysis of collisions and transport at
zero temperature ceases to be valid.

\section{The Kink Mechanism}
\label{review}

A physical mechanism has been proposed for the formation of the layers
which is related to the existence of a ``kink'' \cite{makse,makse2,cms}.
Suppose, e.g., that a pair of static layers is formed with the layer of
larger grains on top of the layer of smaller grains. When an incoming
mass of grains avalanches down the slope, the larger grains reach the
base of the pile first, due to the fact that large grains do not tend to
get trapped (in local minima of the sandpile profile) as easily as small
grains.  Additionally, in the case of rapid flows, the smaller grains
also size segregate to the bottom of the rolling phase due to
percolation so that the larger grains, being at the top of the rolling
phase, tend to travel further since they do not interact with the bulk
phase.

During the avalanche of grains, some small grains are captured in the
static layer of large grains, smoothing the surface and thereby allowing
more small grains to fall downward and eventually reach the bottom of
the pile. When the flow reaches the base of the pile, we see that the
grains develop a profile characterized by a well-defined ``kink'' at
which the grains are stopped. This kink moves in the direction opposite
to the flow of grains, conserving its profile until it reaches the top
of the pile.

In the process of falling down the slope, grains (small and large) stop
at the kink. We see that the smaller grains stop first (since the small
grains are already segregated in the rolling phase) so a pair of layers
forms, with the smaller grains underneath the large grains (see Fig.
\ref{strat}b). When the kink profile reaches the top of the sandpile,
the pair of layers is completed. Then this process is repeated: a new
avalanche occurs, some larger grains reach the bottom of the pile, the
kink is developed, and a new pair of layers is formed.

The size of a pair of layers $\lambda$ is determined by the thickness of
the layer of rolling grains during an avalanche, $R^0$, which in turn is
determined by the flux of adding grains. The volume of rolling grains
$\Omega_{\mbox{\scriptsize aval}}$ that reaches the kink during a time
interval $\Delta t$ and in a differential $dy$ is
\begin{equation}
\Omega_{\mbox{\scriptsize aval}}=\ell \Delta t ~ v(y) dy + 
\ell \Delta t~ v_\uparrow dy,
\label{co}
\end{equation}
where $v(y)$ is the velocity of the rolling grains at a distance $y$
from the pile surface of static grains, $v_\uparrow>0$ is the upward
velocity of the kink which is constant, and $\ell$ is the gap between
the plates of the cell. The first term in (\ref{co}) is the volume of
grains falling down the slope, and the second term represents the volume
of grains from the rolling phase that the kink encounters when it
advances uphill at velocity $v_\uparrow$---i.e., $v(y)+ v_\uparrow$ is
the velocity of the rolling grains in the co-moving frame of the kink.

The volume of grains in a well-formed kink is
\begin{equation}
\Omega_{\mbox{\scriptsize kink}}=\ell \Delta t ~ v_\uparrow dy.
\label{ca}
\end{equation}
Then if all the grains are stopped at the kink, the number of rolling
grains falling down, $\mu_{\mbox{\scriptsize fluid}}
\Omega_{\mbox{\scriptsize aval}}$, where $\mu_{\mbox{\scriptsize fluid}}$ 
is the density of the fluid phase (the number of rolling grains per unit
volume) should scale as the volume of grains in the kink
$\mu_{\mbox{\scriptsize bulk}}\Omega_{\mbox{\scriptsize kink}}$, where
$\mu_{\mbox{\scriptsize bulk}}$ is the density of the bulk phase. Hence
\begin{equation}
\label{conservation0}
 \mu_{\mbox{\scriptsize fluid}}
\ell \Delta t ~ \int_0^{R^0} (v(y) + v_\uparrow) ~ dy 
 = \mu_{\mbox{\scriptsize bulk}}
\ell \Delta t ~ \int_0^\lambda v_\uparrow dy.
\end{equation}
The mean value of the downward velocity of the grains averaged over the
rolling phase is
\begin{equation}
\overline{v} \equiv \frac{1}{R^0} \int_0^{R^0} v(y) dy,
\label{vmeandef}
\end{equation}
so from (\ref{conservation0}) \cite{makse,makse2}
\begin{equation}
\label{conservation}
\lambda =  \frac{\mu_{\mbox{\scriptsize fluid}}}
{\mu_{\mbox{\scriptsize bulk}}} ~\frac{(\overline{v}+v_\uparrow)} {
v_\uparrow}~ R^0.
\end{equation}

The analytical shape of the kink has been obtained in \cite{makse2,cms}.
We introduce four different generalized angles of repose
$\theta_{\alpha\beta}$, corresponding to the interactions between a
rolling grain of type $\alpha$ and a static grain of type $\beta$:

\begin{itemize}
\item{ ~} $\theta_{22}$ corresponds to the angle of repose
of the pure large-cubic species,

\item{ ~} $\theta_{11}$ is the repose angle of the pure small-rounded species
($\theta_{22}>\theta_{11}$), and

\item{ ~} $\theta_{21}$ correspond to the interaction between a 
large cubic rolling grain and small-rounded static grains, and

\item{ ~} $\theta_{12}$ correspond to the interaction between a 
 small-rounded rolling grain and large cubic static grains.
\end{itemize}

For stratification we have \cite{makse2,cms,segre}
\begin{equation}
\label{e.6}
\theta_{21}<\theta_{11}<\theta_{22}<\theta_{12}.
\end{equation}

Since the kink is a traveling wave solution, we can write
\cite{makse2,cms}
\begin{equation}
\label{e.7}
f(x,t)\equiv h(x,t)+\theta_{0} x = f(u),
\end{equation}
where $u\equiv x+v_\uparrow t$, $\theta_0$ is the angle of the pile
after a pair of layers is formed, and $h(x,t)$ is the height of the
static phase (Fig. \ref{stages}c). The solution for the lower layer of
small grains is \cite{makse2}
\begin{equation}
-\frac{1}{w}\log\left(1-\frac{2wf}{R^0}\right) =
\frac{\gamma}{v_\uparrow}(f-\delta_1 u), 
\label{lower}
\end{equation}
where
\begin{equation}
\label{e.10000}
\delta_1 \equiv \theta_0-\theta_{11}>0,
\end{equation}
 $w \equiv v_\uparrow/(\overline{v}+v_\uparrow)$, and $\gamma$ (units of
1/sec) is the rate of collisions between static and rolling grains.
Since we have a pile made of two different species, the angle of the
layers $\theta_0$ is not necessarily either $\theta_{22}$ nor
$\theta_{11}$. However, since the top layer of the stripes is made of
large-cubic grains, the resulting angle $\theta_0$ is closer to
$\theta_{22}$ than to $\theta_{11}$.

The solution for the profile of the upper layer of the kink is
\cite{makse2}
\begin{equation}
\label{upper}
f(u) = \left({R^0\over w}\right) \left[1-\exp \left({w\gamma\delta_2
u\over v_\uparrow}\right) \right],
\end{equation}
where
\begin{equation}
\label{e.10}
\delta_2\equiv\theta_0-\theta_{22}<0. 
\end{equation}
We will compare this solution with the profile of the kink measured
experimentally. Figure
\ref{stages}c shows a sketch of the kink. The angle of the layers is
$\theta_0$, which is an angle between $\theta_{11}$ and $\theta_{22}$,
so that solutions (\ref{lower}) and (\ref{upper}) exist. The lower part
of the kink is made of small-rounded grains and therefore has an angle
close to $\theta_{11}$. At the center of the kink, the larger grains are
captured on top of smaller grains, therefore the angle decreases toward
the cross-angle of repose $\theta_{21}$. Then the angle of the profile
of the kink approaches $\theta_0$.

Dimensional analysis of Eq. (\ref{lower}) indicates that the upward
velocity of the kink is proportional to the flux of arriving grains,
i.e. proportional to $R^0$ \cite{makse2},
\begin{equation}
 v_\uparrow = C_1~ \gamma ~R^0.
\label{vup}
\end{equation}
Here $C_1$ is a numerical constant that does not depend on $\gamma$ or
$R^0$, but may depend on the angles of repose and other features of the
grains.

Below, we will show that the velocity of the rolling grains in an
avalanche increases linearly with the height of the grains in the
rolling phase, implying that the mean value of the velocity $\overline
v$ is proportional to the thickness of the rolling phase. The
coefficient is again proportional to $\gamma$ by dimensional analysis,
\begin{equation}
\overline v = C_2 ~\gamma ~R^0,
\label{vmean}
\end{equation}
where $C_2$ is also a numerical constant that does not depend on
$\gamma$ or $R^0$, but may depend on the angles of repose.

From Eqs. (\ref{conservation}), (\ref{vup}) and (\ref{vmean}) we obtain
the dependence of the wavelength on $R^0$
\begin{equation}
\lambda = \frac{\mu_{\mbox{\scriptsize fluid}}}
{\mu_{\mbox{\scriptsize bulk}}} ~C~R^0,
\label{lambda}
\end{equation}
where $C\equiv 1+C_2/C_1$, is a constant independent of $R^0$.  Thus,
the wavelength increases linearly with the flux of grains. We will test
Eqs. (\ref{lower})-(\ref{lambda}) experimentally.

\begin{figure}
\centerline{
\vspace*{0.5cm}
\vbox{ \epsfxsize=6cm \epsfbox{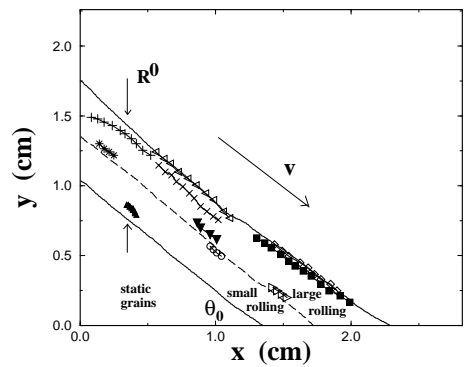} }
    }
\narrowtext
\caption{Trajectories $(x,y)$ 
of the rolling grains during the avalanche for Exp. $\# 5$ in a window
of observation of 2.82 cm by 2.26 cm. The thickness of the rolling layer
is $R^0 = 0.65$ cm. The dashed line is the boundary between the sublayer
of smaller rolling grains and the sublayer of larger rolling grains due to
the percolation effect. The angle of the pile is $\theta_0$.}
\label{trajectories}
\end{figure}

\section{Dynamics of stratification}

To test the above mechanism, we perform a series of six experiments
where we study in detail the dynamics of stratification by measuring all
the quantities involved in the process. We use a Kodak Ektapro 1000
high-speed video camera system to record the motion of the grains. In
order to study the profile of the kink and the effects of percolation in
the rolling phase, we measure the profile of the static and rolling
phases and compare with analytical predictions. We measure the velocity
of the rolling grains, the velocity of the kink, the wavelength of the
layers, and also other phenomenological constants such as $\gamma$, and
all four angles of repose characterizing the mixture
$\theta_{\alpha\beta}$.

According to the picture discussed in Sec. \ref{review}, we study the
dynamical process of stratification by dividing the process in three
stages (Fig. \ref{stages}):

\begin{figure}
\centerline{
\vspace*{0.5cm}
\vbox{ {\bf(a)} \epsfxsize=6cm \epsfbox{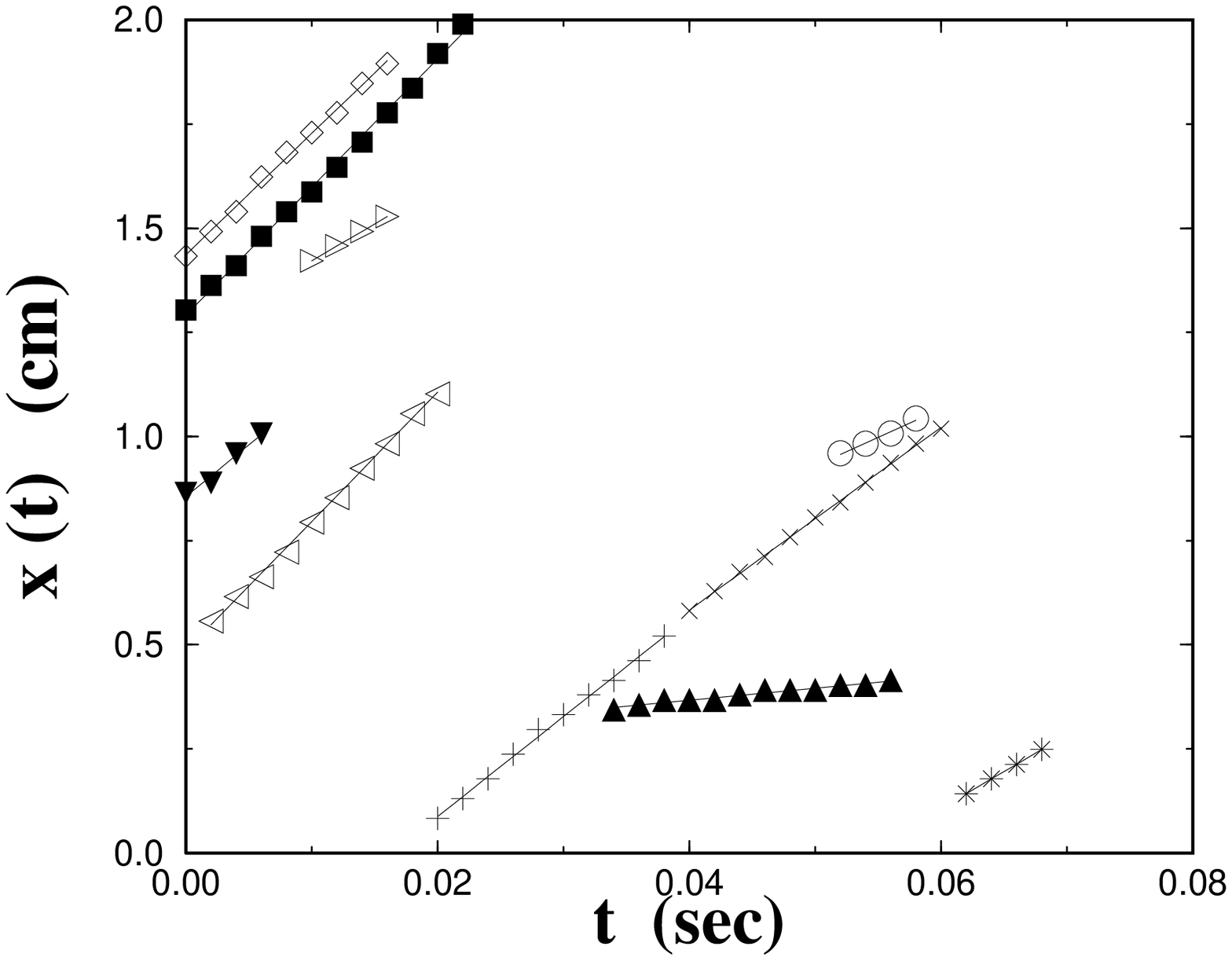} }
    }
\centerline{
\vspace*{0.5cm}
\vbox{ {\bf(b)} \epsfxsize=6cm \epsfbox{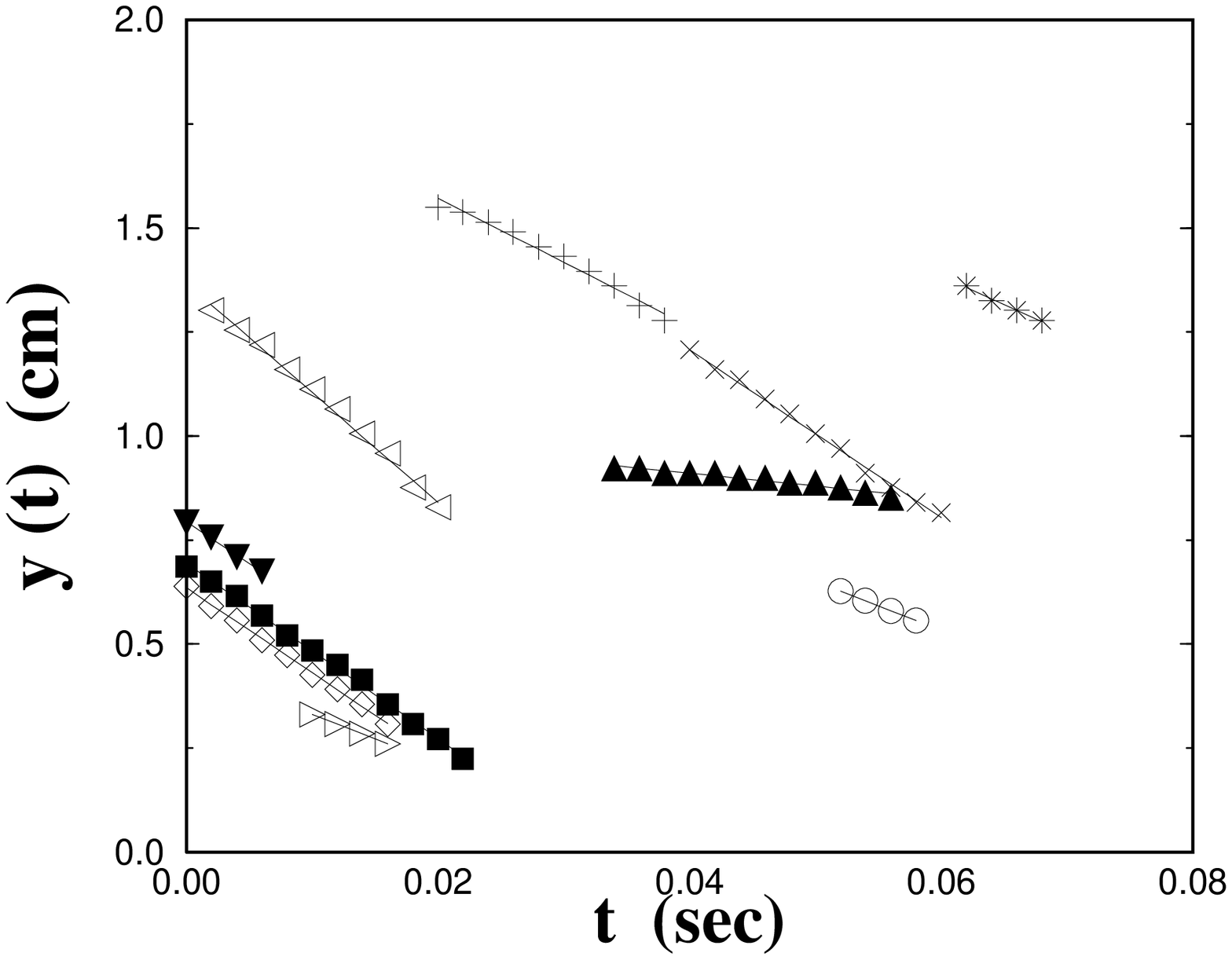} }
    }
\narrowtext
\caption{(a) Position $x(t)$ and 
(b) position $y(t)$ of the rolling grains shown in
Fig.~\protect\ref{trajectories} as a function of time. The straight
lines indicate that the velocities achieve a constant value, given by
the viscous damping and the gravitational driving force. The symbols
correspond to the grains plotted in Fig.
\protect\ref{trajectories}.}
\label{xyt}
\end{figure}

\subsection{Avalanche of grains and percolation effect}

In all six experiments, we use the same mixture and cell, but different
fluxes---i.e., different $R^0$ (see Table \ref{table1}). We focus our
study on a small window of observation of typically 3 cm $\times$ 2 cm
size located in the center of the pile. Using the high speed video
camera at a frame rate of 1000 frames per second, we are able to track
the motion of each individual grain in a downhill avalanche during
1.6\/s of recording time. The rolling grains take only 0.1\/s to cross
the window of observation. However, we continue recording after this
time elapses, so that we can record in the same shot of 1.6\/s the
grains flowing down and the profile of the kink moving up. By tracking
the motion of each individual grain, we are able to measure the velocity
profile of the grains along the layer of moving grains.

The thickness of the layer of rolling grains in all our experiments
ranges from $0.3$ cm to $0.7$ cm. Thus the layer of rolling grains is
thick enough that it is possible to observe the size segregation of the
grains in the rolling phase. For a well-developed flow of grains down an
inclined plane, the grains segregate in the moving layer, with the small
grains at the bottom of the moving layer, and the large grains at the
top. This effect is called 
``free surface segregation'', ``percolation'' or ``kinematic
sieving'' \cite{drahun,drahun2,savage1,savage2,savage3}, and occurs
because the smaller grains percolate downward through the gaps left by
the motion of larger grains in the rolling phase. In our experiments,
the flux of grains is sufficiently high that the layer of rolling grains
is large enough that the percolation effect can be observed in the
rolling phase. Figure \ref{trajectories} shows the trajectories of
several grains for Experiment $\# 5$ (Table \ref{table1}) in a window of
2.82 cm by 2.26 cm, plotted every two frames ($\approx 2$\/ms). We find
that the large grains occupy the top part of the moving layer, and the
small grains are located at the bottom of the moving layer.

\begin{figure}
\centerline{
\vspace*{0.5cm}
\vbox{ \epsfxsize=6cm \epsfbox{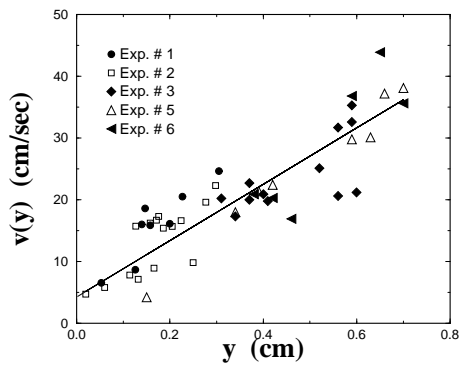} }
    }
\narrowtext
\caption{Velocity parallel to the pile surface
of the rolling grains, corresponding to the experiments listed in
Table~\protect\ref{table1}, as a function of their vertical position $y$
from the top of the static phase. We find a linear velocity profile.}
\label{profile}
\end{figure}

To measure the velocity profile of the grains in the rolling phase, the
position of every rolling grain is spotted on the screen of the video
camera. We follow the trajectory of the grain during a period of time
where the rolling grain is well distinguished from the other grains in
the moving phase. We stop the image every 2\/ms (2 frames) and record
the ($x,y$) position in pixels of the screen. The position thus measured
is manually entered in a data file, giving the ($x,y$) position of the
grain as a function of time. We study the motion of the grains in the
center of the pile, where the grains have achieved a constant velocity
along the direction of the pile surface (the viscous friction force has
balanced the gravitational driving force on the grain). Figure \ref{xyt}
shows the $x$ and $y$ coordinates of the grains of
Fig.~\ref{trajectories} as a function of time. The velocity of the
grains we calculate from the slope of such curves. Indeed, we observe
that the motion of the grains in the center of the pile is overdamped;
the velocity is constant as a function of time.

Due to the percolation effect, the layer of rolling grains is actually
made of two equal size sublayers (since we use an equal volume mixture
of two species) of smaller and larger grains. However, the velocity
profile of the grains is continuous along the thickness of the moving
layer.  Figure \ref{profile} shows the velocity of the grains for the
six experiments listed in Table \ref{table1} as a function of $y$, the
height of the rolling grain to the top of the static phase. The velocity
profile is linear in $y$. Using the data from the experiments listed in
Table \ref{table1} we find
\begin{equation}
v(y) = (46/\mbox{sec}\pm 2)~y.
\label{vy}
\end{equation}
The coefficient of the linear relation (\ref{vy}) is independent of
$R^0$, implying that the mean value of the velocity of the grains
(\ref{vmeandef}) is proportional to $R^0$
\begin{equation}
\overline v = (23/\mbox{sec}\pm 2)~R^0.
\label{vmeanexp}
\end{equation}
By comparing with Eq. (\ref{vmean}) and assuming that the coefficient
$C_1$ is of the order of one, we obtain an estimate of the rate
$\gamma\simeq 23$/sec. Similar velocity profiles have been found in
Ref.~\cite{savage4}, although these results (which were obtained for
single-species grains falling down inclined planes at different angles
above the angle of repose) do not apply to our system, since we are
interested in the velocity profile of the grains avalanching on a
surface composed of large grains at the angle of the layers, as occurs
in the stratification experiment.

\begin{figure}
\centerline{
\vspace*{0.5cm}
\vbox{ \epsfxsize=8cm \epsfbox{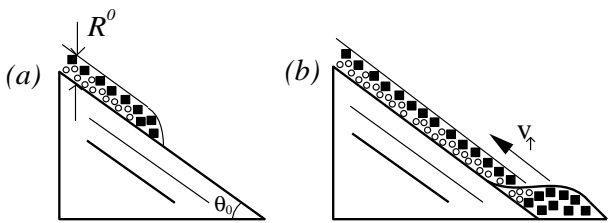} }
    }
\narrowtext
\caption{Formation of the kink at the base of the pile.
(a) The large grains roll down ahead of the smaller grains since they are
at the top of the rolling layer and have larger velocity than the small
grains. (b) When the larger grains reach the substrate they are stopped,
and they act as a wall where the smaller grains are stopped.  The kink
emerges from this interaction.}
\label{base}
\end{figure}

\subsection{Formation of the kink}

The formation of the kink is determined by the interaction of the grains
at the bottom of the pile. When the grains reach the substrate, we find
that the larger grains arrive first and then the smaller grains arrive
(Fig.~\ref{base}a), because the larger grains roll down more easily than
the smaller grains, since they are at the top of the rolling phase due
to percolation. They have larger velocity than the small grains since
they are at the top of the rolling layer (Fig. \ref{base}a). The large
grains are stopped at the substrate, and they develop the profile shown
in Fig.~\ref{base}b.

The condition for the formation of the kink seems to be the existence of
two species, not one. In fact, when we pour single-species grains in a
open cell, we do not observe the stationary kink, but we observe that
the height of the pile and the profile of the rolling phase acquire a
steady state (without oscillations) where the profiles are conserved in
time. In the case of two species, the larger grains reaching the bottom
before the smaller grains act as a ``wall'' or ``incipient kink'' where
the smaller grains are stopped (Fig. \ref{base}b). Thus, when the small
grains arrive near the substrate, they find some large grains already
there.  They are stopped in this way, and the kink emerges from this
interaction. When the kink is developed it starts to move uphill with
constant velocity and conserving its profile.

de Gennes \cite{varenna} has shown that when a flow of single-species
grains flowing down a plane at the angle of repose reaches a {\it
vertical} wall, the grains develop an uphill wave of constant velocity
$v_\uparrow \sim \gamma R^0$. Although this uphill wave is not
stationary as found for the kink in our experiment, the solution found
by de Gennes shows that it is possible for the smaller grains to be
stopped by a moving ``wall'' of large grains and thereby give rise to a
kink. The existence of uphill waves (although not stationary as the kink
solution) was also noticed in Refs.~\cite{bouchaud,bouchaud2}.

\subsection{Uphill motion of the kink and formation of a pair of layers}

By using the video camera at a frame rate of 1000 frames/sec, we can
distinguish the fraction of grains which is in the rolling phase and the
fraction of grains which is strictly immobile, the fundamental
ingredient of the theories of
Refs.~\cite{bouchaud,bouchaud2,pgg,bdg,makse2,cms,makse3}.  Thus we
identify the time behavior of the boundary between the rolling phase and
the static phase, and the profile of the kink.

Since the contrast between the rolling and static grains is not very
good, we must identify this boundary ``manually.'' We play the movie 5
frames back and then 5 frames forward and identify which grains are
moving and which grains are static. We repeat these measurements every
0.05\/s, and record the coordinates of the bulk/fluid interface
\cite{comment}.

\begin{figure}
\centerline{
\vspace*{0.5cm}
 \vbox{ {\bf (a)}
\epsfxsize=6cm \epsfbox{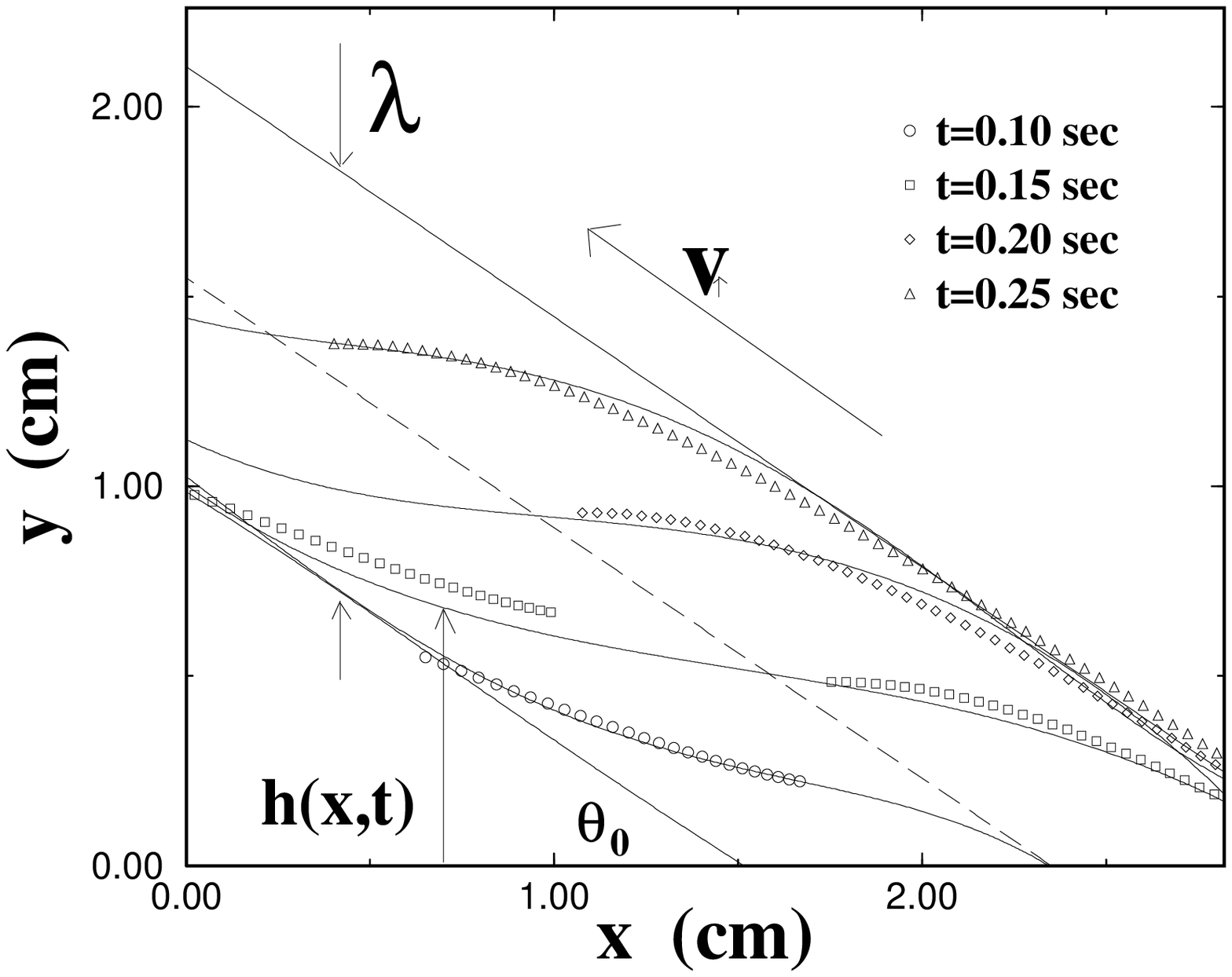} }
    }
\vspace*{0.5cm}
\centerline{
\vspace*{0.5cm}
\vbox{ {\bf (b)}
\epsfxsize=6cm \epsfbox{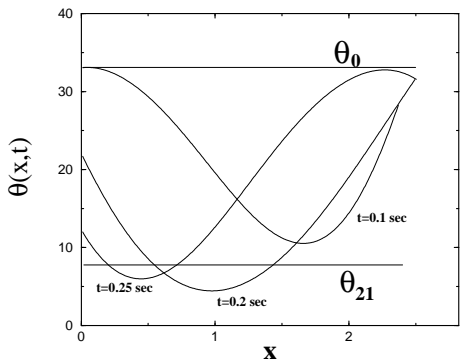} }
    }
\narrowtext
\caption{{\bf (a)}
Profile of the kink $h(x,t)$ obtained in Experiment \#5, shown at time
intervals of 50\/ms. The dotted line is the boundary between the static
layers of smaller and larger grains. The kink moves uphill with constant
velocity $v_\uparrow$. The symbols correspond to the fits to the
analytical solution (\protect\ref{lower}) and (\protect\ref{upper}).
The data are first obtained manually and the fitted with a polynomial.
{\bf (b)} Angular profile of the kink obtained from Fig. (a),
$\theta(x,t) = -\mbox{arctan} (\partial h(x,t)/\partial x)$. The angle
is between $\theta_0$ and $\theta_{21}$ as explained in Fig.
\protect\ref{stages}c. }
\label{kink}
\end{figure}

Figure~\ref{kink}a shows the profile of the kink as a function of $x$
plotted every 0.05 sec in a window of 2.82 cm by 2.26 cm (Experiment
\#5), and one sees the profile of the kink moving upward with constant
velocity $v_\uparrow$. Notice that the profile of the kink is
stationary. Figure \ref{kink}b shows the angle of the kink profile
\begin{equation}
\label{e.13}
\theta(x,t)\equiv - \mbox{arctan} (\partial h(x,t)/\partial x),
\end{equation}
where $h(x,t)$ is the profile of the kink obtained from Fig.
\ref{kink}a, as a function of $x$ measured at different times. We fit
the analytical solution of the shape of the kink (\ref{lower}),
(\ref{upper}) obtained in \cite{makse2} to the experimental profile of
the kink and find good agreement (Fig. \ref{kink}a). We find the best
fit for values $\gamma = 150$/sec, $\overline{v} = 15.6$ cm/sec,
$v_\uparrow = 16.2$ cm/sec, $R^0 = 0.7$ cm, $\theta_0 = 33^{o},$
$\theta_{11} = 25^{o}$, and $\theta_{22} = 43^{o}$.

The solution (\ref{lower}) and (\ref{upper}) is valid only for the lower
and upper part of the kink, so the center of the kink, where small and
large grains are mixed, cannot be reproduced. The values we use to fit
the analytical solution are somehow different from the values we
measure. However, we note that the exact shape of the kink depends on
the type of interaction term used in the equation of motion to describe
the rolling/static grains interaction. In particular, the interaction
term used in \cite{makse2} does not include nonlinear terms that might
be important when the flux of grains becomes large (we comment on this
point in Sec.~\ref{results}). However, the fair agreement between
experiment and theory indicates that some features of the interaction
have been captured by this approach.

We also focus on the different collision processes between rolling
grains in contact with the solid surface and the static grains. We find
that amplification process (i.e. when a rolling grain removes a static
grain via a collision) do not happen very often. The percolation effect
forbids the larger grains to interact with the surface, thus prohibiting
cross-amplification processes of the type of a larger rolling grain
amplifying a smaller static grain. The main interaction seems to be the
capture of rolling grains at the surface---when a rolling grain is
converted to the static phase. However, we emphasize that it is
difficult to clearly determine the nature of the interaction at the
surface (capture versus amplification) because the smaller grains are
the only interacting grains in the region of observation and they are
difficult to track.

\begin{figure}
\centerline{
\vspace*{0.5cm}
\vbox{ {\bf (a)}\epsfxsize=6cm \epsfbox{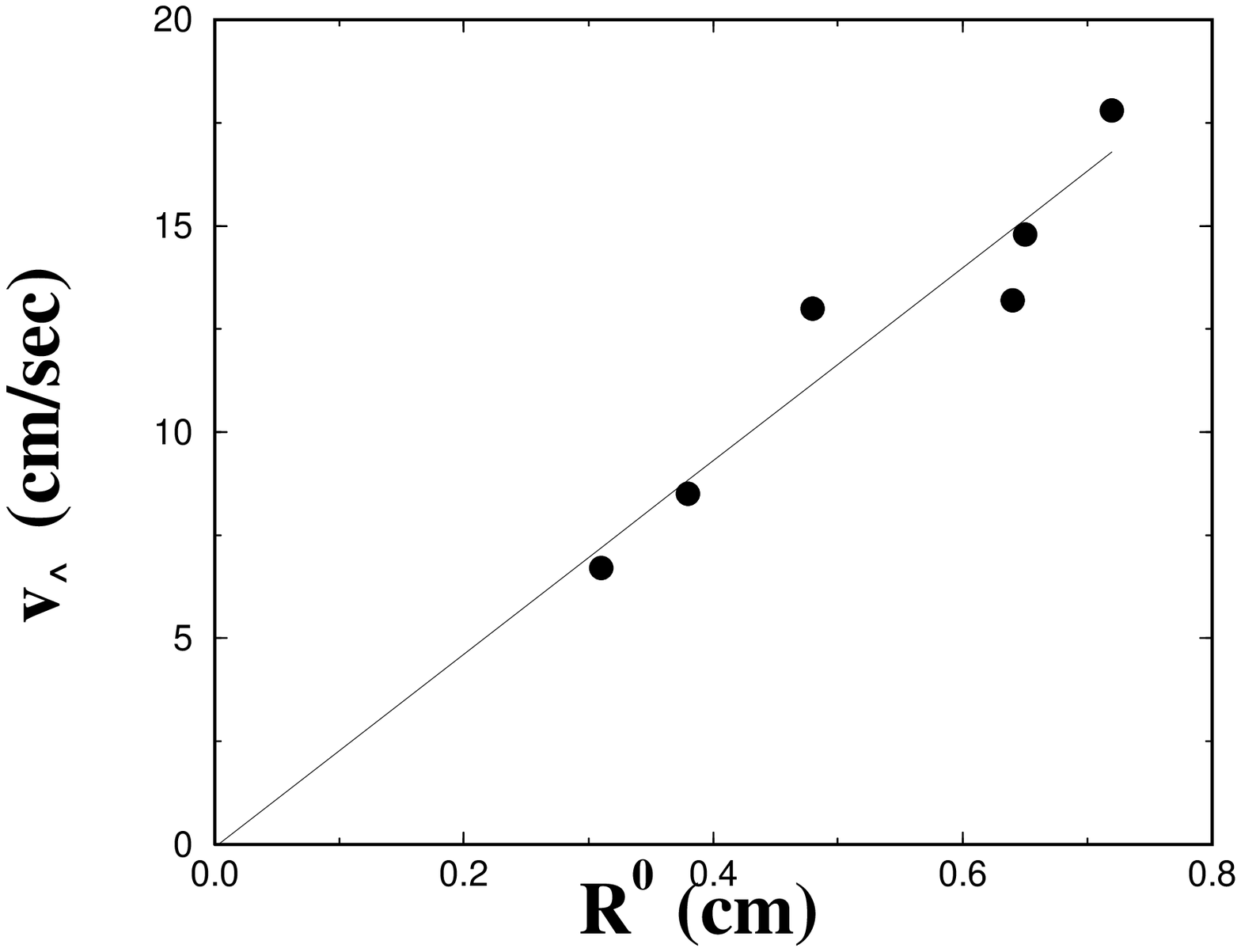} }
    }
\vspace*{0.5cm}
\centerline{
\vbox{{\bf (b)} \epsfxsize=6cm \epsfbox{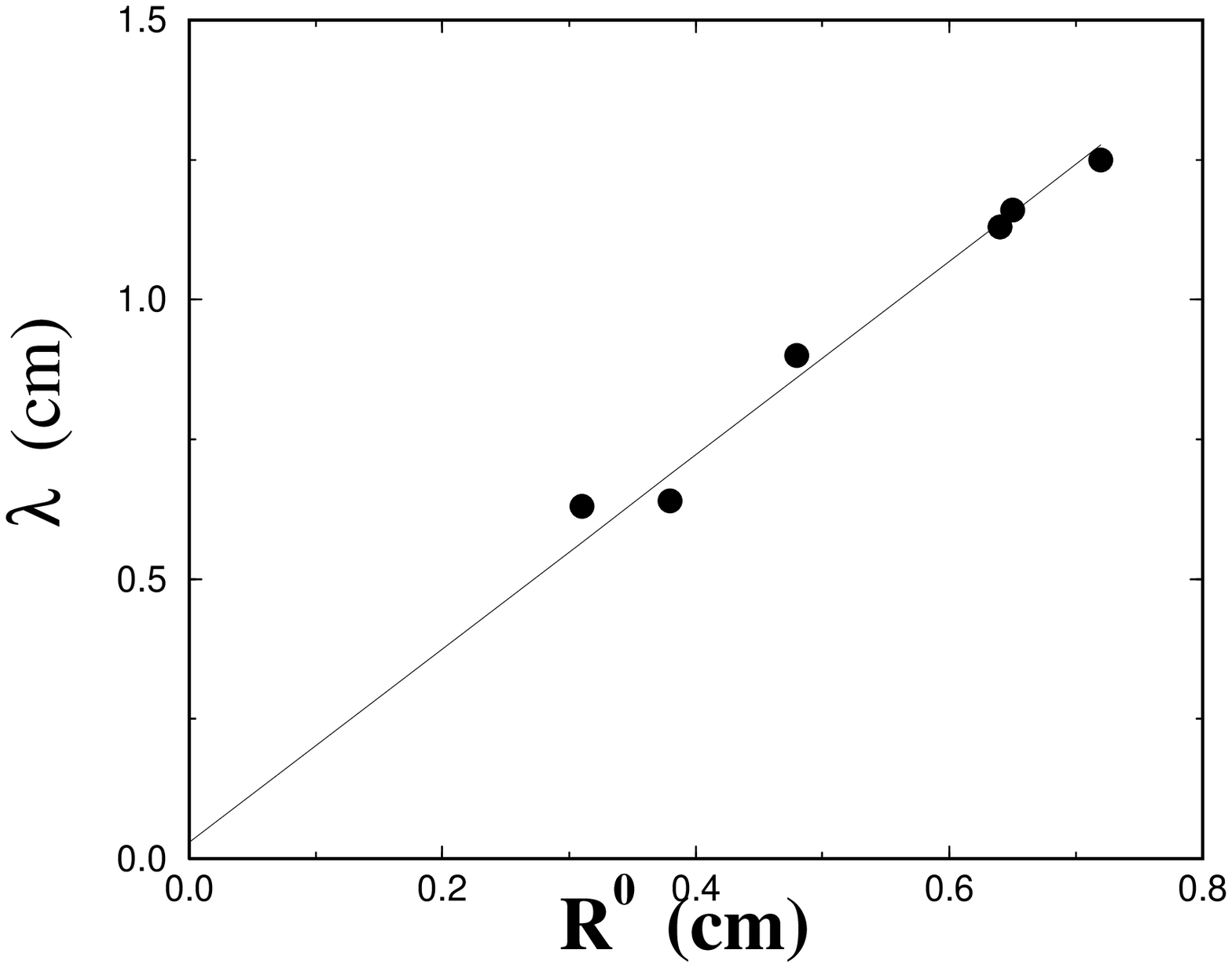} }
    }
\narrowtext
\caption{{\bf (a)}
Velocity of the kink $v_\uparrow$ as a function of the thickness of the
rolling layer $R^0$ for the six experiments (Table
\protect\ref{table1}). {\bf (b)} Wavelength of the layers as a function
of the thickness of the rolling layer $R^0$ for the experiments  (Table
\protect\ref{table1}). }
\label{r0-vup}
\end{figure}

We measure the velocity of the kink $v_\uparrow$ as a function of $R^0$
(see Table \ref{table1}). Figure \ref{r0-vup}a shows the results which
can be fit to a straight line. We find
\begin{equation}
v_\uparrow = (23\mbox{/sec}\pm2)~ R^0,
\label{vupexp}
\end{equation}
where $R^0$ is typically 5~cm. The velocity of the kink is approximately
the same as the mean velocity of the rolling grains ($C_2 \simeq C_1$).
Comparing with Eq. (\ref{vup}), we obtain a second estimate of the rate
$\gamma \simeq 23$/sec.

Figure \ref{r0-vup}b shows the wavelength of the layers as a function of
the thickness of the rolling phase for the six experiments of Table
\ref{table1}. The data can be fit to a straight line, and we find
\begin{equation}
\lambda = (1.7\pm0.1) ~ R^0,
\label{lambdaexp}
\end{equation} 
which agrees with the prediction of Eq.~(\ref{lambda}). Comparing with
Eq.~(\ref{lambda}), we obtain $(\mu_{\mbox{\scriptsize
fluid}}/\mu_{\mbox{\scriptsize bulk}}) C\simeq 1.7$. Using
(\ref{vmeanexp}) and (\ref{vupexp}) we obtain $C \simeq 2$, so that $
\mu_{\mbox{\scriptsize fluid}} / \mu_{\mbox{\scriptsize bulk}}
\simeq 0.85$, corresponding to the fact that the fluid phase is 
less dense than the bulk.

\section{Phenomenology}
\label{results}

Table \ref{table2} shows the values of the phenomenological constants
measured for the equal volume mixture of quasi-spherical glass beads of
mean diameter $0.19$ mm and cubic-shaped sugar grains of typical size
$0.8$ mm. The mean value of the velocity of the grains falling down the
slope and the velocity of the kink range from $7$ cm/sec to $17$ cm/sec.
As noted above, $\gamma\approx 23$/sec and $R^0\approx 0.5$ cm.

According to Eq. (\ref{lower}), the lower part of the kink has the angle
of the small-rounded grains $\theta_{11}$, and near the layer of large
grains there are logarithmic corrections. The angle of the kink
decreases towards the center of the kink; it is equal to $\theta_{21}$
at the position where the larger grains start to be captured
(Fig.~\ref{stages}c) Thus, measuring the minimum of the angle of the
profile of the kink at the transition from the layer of smaller grains
to the layer of larger grains provides a method to measure the value of
the crossover angle of repose $\theta_{21}$. Then, assuming that
$\theta_{11} - \theta_{21} = \theta_{22} - \theta_{12}$ \cite{makse2},
the critical angle $\theta_{12}$ can be obtained too.

We measured for the angle of repose of the pure species
\begin{equation}
\label{e.15}
\theta_{11} = 26^\circ \pm 1,\qquad \theta_{22} = 39^\circ\pm 1.
\end{equation}
Figure \ref{kink}b shows the angle of the pile near the kink at
different times. From these curves we measure the angle of the layers
\begin{equation}
\theta_0 \simeq 33^{o},
\end{equation}
and the remaining angle of repose of larger grains on smaller grains

\begin{equation}
\theta_{21} \simeq 8^{o}, ~~~~~~~\mbox{and} ~~~~~~~
\theta_{12} \simeq 57^{o}. 
\end{equation}

The constant $\overline{v}/\gamma$ represents the distance at which a
rolling grain stops in a pile at an angle different from the angle of
repose \cite{pgg}---i.e., $\overline{v}/\gamma$ represents the distance
at which a rolling grain is stopped at the kink. The constant
$\overline{v}/\gamma \simeq 0.3-0.7$ cm is small compared to the system
size $L=30$ cm---i.e., $\overline{v}/\gamma$ scales with the size of the
grains and not with $L$, as expected \cite{pgg}. Notice also that
$\overline{v}/(\gamma \tan\psi) \simeq 2$cm---where $\tan\psi\equiv\tan
\theta_{11} - \tan \theta_{21} \simeq 0.3$---is the size of the region
of mixing in the center of the pile observed in the case of segregation
of the mixtures of smaller cubic grains and large rounded grains
(Fig.~\ref{strat}b) \cite{makse3}. This region of the mixture is
observed to be small in comparison with the systems size.

Finally, we comment on the application of the theoretical calculation of
the model developed in \cite{makse2,cms,makse3} to the problem of
stratification when percolation effects are acting. The dependence of
the repose angle on the composition of the surface proposed in
\cite{makse2,cms,makse3} is analogous to the effect
of percolation. Due to percolation, only the smaller
grains interact with the surface, causing the larger ones to be
convected further. The models of
\cite{makse2,cms} use the fact that the repose angle of the smaller grains
is always larger than the repose angle of the larger grains for a given
composition of the surface (i.e., $\theta_{11}>\theta_{21}$, and
$\theta_{12}>\theta_{22}$), then the smaller grains are always the first
to be trapped, and the large ones are always convected down as it occurs
due to the percolation effect. Moreover, capture of larger grains is
forbidden on top of smaller grains since the capture function of the
large grains is zero around the angle of repose of the small grains
\cite{makse3}.

A simple extension to explicitly include percolation effects in the
formalism of \cite{makse2,cms} shows only small corrections to the
profiles of the rolling and static grains, which provides evidence for
the possible applicability of the results of \cite{makse2,cms} to the
case where percolation effects take place in the rolling phase. However,
caution must be taken in the definition of the fluid/bulk interaction in
the theoretical formalism. We take the interaction term to be
proportional to the thickness of the rolling phase, 
an approximation valid for thin flows
\cite{varenna,pgg}.  Although this approximation might be still valid in
the case of thick flows \cite{cms}--- since the interaction might be
proportional to the preassure exerted by the fluid phase \cite{zik},
which in turn is proportional to $R^0$ for a fluid--- nonlinear terms
might be also necessary to completely describe the flow, especially
because the interaction among the rolling grains (which is neglected in
the theoretical formalism) becomes important.

The dependence of the velocity of the rolling grains on the position of
the grain in the rolling phase is another fact not included in the
theory, which considers a uniform velocity for all the grains in the
rolling phase.  For a comparison with the theory, we have replaced the
velocity of the grains in the theory by the mean value of the velocity
of the rolling grains measured experimentally.

\section{Other Limits}

We notice that the dependence on the plate separation $\ell$, although
present in Eqs. (\ref{co}) and (\ref{ca}), has disappeared in
Eq.~(\ref{lambda}), and the relevant length scale that determines the
wavelength is $R^0$. However, a change in the flux of adding grains, or
a change in the gap $\ell$, changes the wavelength since $R^0 \sim
\mbox{flux}/\ell$.  Thus, e.g., by keeping the flux constant and
increasing the gap $\ell$, we find a decrease in $R^0$, and we find that
the wavelength $\lambda$ decreases according to (\ref{lambda})
\begin{equation}
\label{e.11}
\lambda\sim 1/\ell.
\end{equation}
This dependence has been measured in \cite{kaka2,herrmann3}.

For the moderate fluxes used in this study (of the order of 1 gr/sec),
the role of the flow rate is to determine the wavelength according to
Eq. (\ref{lambda}). For larger fluxes, Eq. (\ref{conservation}) is still
valid as long as the kink mechanism works. However, deviations from the
linear dependence of Eq. (\ref{lambda}) might occur since the velocity
of the kink and the velocity profile of the rolling grains might deviate
from the linear regime. The densities of the bulk and fluid phase might
also change with the flux of grains, giving rise to nonlinear relation
between $\lambda$ and $R^0$.  For sufficiently large flow rates, the
kink mechanism required to form layers cannot occur (especially the
appearance of the first kink at the onset of the instability
\cite{makse3}), since the grains acquire large momentum, and avalanches
that terminate by an upward moving kink before the next avalanche begins
cannot occur.  In this case, the kink is not able to stop the arriving
rolling grains anymore; the grains ride over the kink so that no
segregation at the kink is possible.  Therefore, for this limit, the
stratification pattern disappears when the flux is sufficiently 
large. Such effect was recently observed in   \cite{baxter} where the 
flux was increased by
factor of 100.

Another deviation from Eq. (\ref{lambda}) might occur at very low flow
rate.  Here the percolation effect disappears and the grains segregate
due to size, because the larger grains do not find large enough holes in
the surface so they roll easier than smaller grains.  In this case, the
rolling phase is thin, so that it behaves as a homogeneous phase with a
constant velocity $v$ for all the grains in the fluid phase.  In this
case, from Eq.~(\ref{conservation}) we obtain
\begin{equation}
\lambda =  \frac{\mu_{\mbox{\scriptsize fluid}}}
{\mu_{\mbox{\scriptsize bulk}}} ~\left(\frac{v} {C_1 \gamma} +
R^0\right).
\end{equation}
Thus when $R^0 \to 0$
\begin{equation}
\lambda \to v/(C_1 \gamma)\sim d,
\end{equation} 
where $d$ is a small constant of the order of the grain size. This lower
limit might indicate the crossover from a percolation regime to a
non-percolation regime at low flow rates.
  
\section{Discussion}

In summary, we tested experimentally the main assumptions of the theory
of surface flows of granular materials.
We measured the profile of the static and rolling phases, in order to
study the profile of the kink and the effects of percolation in the
rolling phase and compared with analytical predictions. We characterized
the dynamical process of stratification by measuring all the relevant
quantities. We measured the velocity of the rolling grains, the velocity
of the kink, the wavelength of the layers, and also the rate of
collisions $\gamma$, and all four angles of repose
$\theta_{\alpha\beta}$ characterizing the mixture. The velocity of the
kink and the wavelength of the layers both vary linearly with the grain
flux. The velocity profile of the grains in the rolling phase is a
linear function of the position of the grains along the moving layer,
which implies a linear relation between the mean velocity and the
thickness of the rolling phase. We find the mean velocity of the rolling
grains is the same as the velocity of the kink.

Our results apply to the
moderate flow rates used in this work of the order of 1 gr/sec.  For
sufficiently larger or smaller flow rates, deviations might appear as
discussed in the text.
For larger fluxes, nonlinear terms may modify Eq. (\ref{lambda}).
For even larger fluxes the kink may not support the 
incoming grains turning  stratification 
 into the mixing of grains or to weak segregation.
For smaller fluxes than the ones used in this study, 
the percolation effect does not take place. 
However, when the size ratio is large enough--- $\rho>1.5$---
strong segregation occurs anyway at the shear surface between the fluid 
and solid phase:
the large grains are not 
trapped in the holes of the surface, and they are convected further.
Thus stratification is also observed  for small 
fluxes, but the size segregation
mechanism is different from the one studied here.
The sharp segregation 
profiles with a separation zone of a few cm 
observed in the experiment shown in Fig. \ref{strat}b is also 
a consequence of 
strong segregation effects.
When   $\rho<1.5$, size segregation has a weak effect, resulting
 in a weak continuos 
segregation of the mixture no matter the shapes of 
the grains. 
Theoretically the case $\rho<1.5$ 
is treated in \cite{bmdg}, and the case $\rho>1.5$ in \cite{makse2,makse3,cms}.

Our results might be also applicable to other systems. Size and 
shape segregation in rotating drums may be  analized
 in analogy to the regimes found
here. Further experimental results may include qualitative studies of the
other flows regimes mentioned above. It would be also appropriate to have an
estimation of the angles of repose of the grains independently of the
theoretical calculations used here. For instance, by gluing grains 
of one species to an inclined plane and pouring
 grains of the other species is a 
way to obtain a direct estimation of the cross angles of repose.

ACKNOWLEDGEMENTS. We thank T. Boutreux, P. Cizeau,  S.
Havlin, H. Herrmann, 
P. R. King, and S. Tomassone for stimulating discussions. HAM
and HES acknowledge support from BP and NSF.

\begin{table}
\narrowtext
\caption{Results of the six experiments.}
\begin{tabular}{|c|c|c|c|c|}
Experiment \#& $R^0$ (cm) & $\lambda$ (cm) & $v_\uparrow$
(cm/sec)&$\overline v$ (cm/sec) \\
\tableline 
1& 0.31 & 0.63 & 6.7& 7.1\\ 2& 0.38 & 0.64 & 8.5& 8.7\\ 3& 0.48 & 0.90 &
13.0 & 11.0\\ 4& 0.64 & 1.13 & 13.2 & 14.7\\ 5& 0.65 & 1.16 & 14.8 &
14.9\\ 6& 0.72 & 1.25 & 17.8 & 16.6\\
\end{tabular}
\label{table1}
\end{table}

\begin{table}
\narrowtext
\caption{ Typical values of phenomenological constants.}
\begin{tabular}{|c|c|c|c|c|c|c|c|}
$\gamma$ (1/sec) & $\theta_{21}$ & $\theta_{11}$ & $\theta_{22}$ &
$\theta_{12}$ & $ v_\uparrow $ (cm/sec) & $\overline v$ (cm/sec) &
$\overline{v}/\gamma$ (cm) \\
\tableline
23& $8^\circ$ &$26^\circ$ &$39^\circ$ &$57^\circ$ & 7-18 & 7-17 & 0.3-0.8 \\
\end{tabular}
\label{table2}
\end{table}


\end{multicols}


\begin{references}
\bibitem{segre}
For the segregation observed in Fig. \protect\ref{strat}b, we have
$\theta_{21}<\theta_{22}<\theta_{11}<\theta_{12}$
\protect\cite{makse2,cms}.
\bibitem{comment}
Since we are considering the limit of high flux, the percolation effect
takes place, so the interaction between rolling and static grains (at
least in the center of the pile where we measure the profiles) is almost
entirely between small grains. This fact makes difficult the
identification of the boundary since small grains are difficult to track
at the resolution of the video camera. Nevertheless we are able to
identify the two phases.
\end{references}
\end{document}